\documentclass[twocolumn,secnumarabic,amssymb, nobibnotes, aps, prd,
 superscriptaddress]{revtex4}
\pdfoutput=1

\usepackage[utf8]{inputenc}
\usepackage{graphics,graphicx}
\usepackage{amsmath,amssymb,amsfonts,latexsym,cancel}
\usepackage{bm}

\begin{document}

\title{Polytropic spheres with electric charge: compact stars,
the Oppenheimer-Volkoff and Buchdahl limits,  and
quasiblack holes}

\author{Jos\'e D. V. Arba\~nil}\email{jose.arbanil@ufabc.edu.br}
\affiliation{Centro
de Ci\^encias Naturais e Humanas, Universidade Federal do ABC, Rua
Santa Ad\'elia 166, 09210-170 Santo Andr\'e, SP, Brazil.}
\author{Jos\'e P. S. Lemos}\email{joselemos@ist.utl.pt}
\affiliation{Centro Multidisciplinar de Astrof\'isica - CENTRA,
Departamento de F\'isica, Instituto Superior T\'ecnico - IST,
Universidade de Lisboa - UL, Avenida Rovisco Pais 1, 1049-001
Lisboa, Portugal.}
\author{Vilson T. Zanchin}\email{zanchin@ufabc.edu.br}
\affiliation{Centro
de Ci\^encias Naturais e Humanas, Universidade Federal do ABC, Rua
Santa Ad\'elia 166, 09210-170 Santo Andr\'e, SP, Brazil.}

\begin{abstract}
We explore a class of compact charged spheres made of a charged
perfect fluid with a polytropic equation of state. The charge density
is chosen to be proportional to the energy density. The study is
performed by solving the Tolman-Oppenheimer-Volkoff (TOV) equation
which describes the hydrostatic equilibrium. We show the dependence
of the structure of the spheres for several characteristic values of
the polytropic exponent and for different values of the charge
densities. We also study other physical properties of the charged
spheres, such as the total mass, total charge, radius and sound speed
and their dependence on the polytropic exponent. We find that 
for the polytropic exponent $\gamma=4/3$ the Chandrasekhar 
mass limit coincides with the Oppenheimer-Volkoff
mass limit. We test the
Oppenheimer-Volkoff limit for such compact objects. We also analyze
the Buchdahl limit for these charged polytropic spheres, which happens
in the limit of very high polytropic exponents, i.e., for a stiff
equation of state. It is found that this limit is extremal and it is a
quasiblack hole.

\end{abstract}
 \maketitle

\section {Introduction}
\label{sec-introd}

\subsection {Compact stars and the TOV equation and method}
\label{sec-tov}

\subsubsection {Compact stars and the Oppenheimer-Volkoff limit}
\label{sec-chandra}

Eddington, in discussing the internal constitution of stars, understood that
by carefully choosing the temperature distribution along a given star, the
gas could be made to obey a polytropic relation between its hydrostatic
pressure $p$ and mass density $\rho$, namely, $p = \omega \rho^\gamma$, where
$\omega$ and $\gamma$ are replaceable constants \cite{ed}. Following the
tradition of calling spherically symmetric star models as gas spheres, he
called these particular models polytropic gas spheres. The stars in question
are stars like the sun, supported against gravitational collapse by matter
and radiation pressure. By obtaining a consistent picture, his initial model
for the stars was vindicated. On the other hand, when he and others tried to
discuss the structure of white dwarfs, stars with much higher density and
compact, they got it all wrong.

Chandrasekhar \cite{chandra1,chandra3} showed then definitely that white
dwarfs are compact stars in which the pressure support against collapse comes
from the quantum degeneracy of the electrons. The temperature throughout the
star is negligible for its structure, and thus these stars, besides being
compact, are also cold. In turn, this means there is an effective simple
polytropic equation between pressure and density. Moreover, as the
configuration gets more compact, the electrons get more and more relativistic
and the last equilibrium configuration is a star with a definite mass shrank
to zero radius. This mass is called the Chandrasekhar limit, generally taken
as $1.44\,M_\odot$ \cite{chandra1,chandra3}.

Now, any given mass in zero radius should be treated in the context of
general relativity. This was done by  Tolman \cite{tolman} and Oppenheimer
and Volkoff \cite{volkoff} who showed that equilibrium configurations with
neutrons, the neutron stars, much more compact than white dwarfs, could also
be achieved, and that these neutron stars have again a mass limit, called the
Oppenheimer-Volkoff limit. Whereas the Chandrasekhar limit appears within
Newtonian gravitation when joined to relativistic kinematic effects in the
degenerate matter, the Oppenheimer-Volkoff limit is a pure general
relativistic limit. It appears because the pressure that supports the star
also has an energy associated with it. When the pressure is sufficiently
large, as for instance in a very compact neutron star, the pressure
contribution to mass-energy makes the gravitational field large enough that
it cannot be supported by the pressure itself. From then on, the star is
unstable to collapse, giving the Oppenheimer-Volkoff limit. Through heuristic
arguments, Landau had also found the Chandrasekhar and the
Oppenheimer-Volkoff limit \cite{landau}.

The question to what happen to stars that get more massive, or more
compact, than allowed by the Oppenheimer-Volkoff limit was answered by
Oppenheimer and Snyder \cite{snyder} who showed that totally collapsed
configurations, black holes, form.  These are objects with a central
singularity that somehow contains what was the star's matter, and with
an event horizon from inside which nothing can escape.

The basic theory for white dwarfs, neutron stars and black holes was
thus laid down in the decade of 1930.

\subsubsection {The TOV equation and method and 
the structure
of compact stars}
\label{sec-tovandstruct}

In their work, Tolman \cite{tolman} and specially Oppenheimer and Volkoff
\cite{volkoff} devised a method to find solutions in a consistent
manner, most prone to numerical integration. They managed to put the
structure equation for cold matter, i.e., matter with zero or negligible
temperature, as ${dp}/{dr}=-(p+\rho){\left(4\pi pr+{m}/{r^{2}}
\right)}/{\left(1-{2m}/{r} \right)} $, where $p=p(r)$ is the pressure as a
function of the radius $r$, $\rho(r)$ is the energy density, $m(r)$ is the
mass
function given by the equation $dm/dr= 4\pi\,r^2\rho$. This equation for
the pressure is the Tolman-Oppenheimer-Volkoff (TOV) equation. In order to
find $p(r)$ an equation of state of the form $p(r)=p(\rho(r))$ should be
provided. One then gives the value zero for the central mass ($m(r=0)=0$) so
that there are no singularities beforehand and some adequate value for the
central pressure. At the point along the radial direction where $p=0$ one
finds the surface of the star, with radius $R$, and with a given total mass
$M$. This solution can then be joined into the correct exterior vacuum
solution, the Schwarzschild solution, giving a full description of the star's
spacetime.  For a consistent procedure see, e.g., \cite{misneresharp,misner}.
For an explanation of the possible different equations of state see
\cite{htww} (see also \cite{thornereview} which in addition gives a general
review on relativistic star structure). A simple equation of state is the
polytropic one $p=\omega \rho^\gamma$, where $\omega$ is a constant and
$\gamma$ is the polytropic exponent, as proposed by Tooper \cite{tooper1} in
a general relativity setting, who also found solutions for general
relativistic compact stars. For instance the interior Schwarzschild solution
has $\gamma\to\infty$ and so the equation of state for the matter is
$\rho={\rm constant}$, with the pressure $p(r)$ adjusting itself to yield a
static configuration.  In \cite{misner} the TOV method is used to find the
exact interior Schwarzschild solution by imposing the equation of state
$\rho={\rm constant}$ in the TOV equation.

\subsubsection {The Buchdahl limit}
\label{sec-buchsimple}

There is also an interesting and important limit in the theory of
compact stars. It is the Buchdahl limit \cite{buchdahl}, and it is a
limit of limits. The Buchdahl limit establishes that for a perfect
fluid sphere of radius $R$ and mass $M$, if $R\geq\frac94\,M$ (or
$R\geq\frac98\,r_+$, where $r_+=2M$ is the gravitational radius) then
there is no equilibrium solution whatsoever. It means that,
independently of the equation of state, this limit is absolute. It can be
found by searching for the condition that yields an infinite pressure
at the star's center. For instance, the Schwarzschild interior
solution, with constant energy density for the matter (the stiffest
equation of state that one can imagine), is a concrete example of the
Buchdahl inequality.  Presumably, when the limit is violated, the
object collapses into a black hole.

Note the difference between the Oppenheimer-Volkoff limit and the
Buchdahl limit. The Oppenheimer-Volkoff limit operates on a set of
perfect fluid configurations whose matter obeys a given equation of
state, say $p=\omega \rho^\gamma$, with $\omega$ and $\gamma$ fixed,
but whose central energy density is increased from member to member of the
set. It gives a maximum mass, and through the radius-mass relation, a
minimum radius. For a neutron star the maximum mass is of the order of
$M\simeq 3\, M_\odot$ and the corresponding minimum radius is $R\simeq
1.7\, r_+$. The Buchdahl limit operates on a set of perfect fluid
configurations, whose members have the stiffest equation of state
$\rho={\rm constant}$, for any constant, and whose central pressure is
increased from member to member of the set up to infinity. This yields
the limit $R\geq\frac98\,r_+=1.125\, r_+$.

\subsubsection {In brief}
\label{sec-brief}

Compact object is a term used to refer to astrophysical objects whose
nature is related to the existence of pressure degeneracy to sustain
the object against gravitational collapse and have in common the
feature that they are all small for their mass.  White dwarfs
\cite{chandrabook}, neutron stars \cite{glendenning}, and black holes
\cite{novikzel,mtw,bron} are well documented and well known objects, see also
\cite{israel}.

Two important quantities in the study of compact objects are the
Oppenheimer-Volkoff mass limit, and the Buchdahl radius to mass limit.

\subsection {Electric compact stars}
\label{sec-electrictov}

\subsubsection{Electric compact stars, the electric
TOV equation and the electric Oppenheimer-Volkoff limit}
\label{sec-eltov}

Given that compact objects exist, a natural question that
can be made is what is the maximum compactness that such an 
object can stand. Matter fields of perfect fluid type
with a stiff equation of state cannot get more compact 
than the Buchdahl limit. Is it the case that allowing for other
kinds of fields or even modifying somehow the gravitational 
field into some extension of general relativity one can 
get frozen stars, i.e., 
stars as compact as their own gravitational radii?

The simplest known field that can be added to a given matter within
the star is the electric field.  By finding a generalized TOV equation
\cite{bekenstein}, one can seek for solutions of electric charged
compact objects. Imposing different equations of state some of these
solutions have indeed been devised
\cite{zhang,defelice_yu,defelice_siming,ray,ghezzi2005,siffert}.

Among the several new properties of these electric 
compact objects it was found that the
Oppenheimer-Volkoff limit, the mass limit that sets in when the 
gravitational field due to the pressure overcomes its own 
support, gets larger as the electric charge of the matter
increases, with the radius of the limiting configuration 
also increasing, see
e.g. \cite{ray}.

\subsubsection {The electric Buchdahl limit}
\label{sec-buchquasi}

The Buchdahl limit also gets modified when electric charged is added to
the matter particles. For charged spheres, the analogous of the Buchdahl
limit has been worked out first in \cite{yu}. A development was
performed in \cite{mak} and in \cite{andreasson_charged} where a sharp
bound for the star's radius $R$ to 
mass $M$ quotient, $R/M$ is given,
namely, ${R}/{M}\geq\left({1}/{3}+\sqrt{{1}/{9}+{Q^{2}}/{3R^{2} }}
\right)^{-2}$. Here it was considered that the matter obeys the 
conditions $p+2p_{T}\leq\rho$, $p\geq0$ and $\rho\geq 0$, where $\rho$,
$p$, and $p_{T}$ are the energy density, the radial pressure and the
tangential pressure of the fluid, respectively. Interestingly, for $Q=M$
this result admits the extremal case $R=M$ solution. Since in this
extremal case the horizon radius $r_+$ is $r_+=M$, one finds that for a
star satisfying the extremal condition $Q = M$ the Buchdahl limit is
such that the radius of the star is as compact as the gravitational
radius $R=r_+$. 

This is a remarkable result in many ways. Are there star
solutions that are as compact as their own gravitational radii?
Certainly there are.

\subsubsection {Electric compact stars and other methods: The Buchdahl
and the quasiblack hole limits}
\label{sec-elotherm}

The TOV equation is a synthetic method to encode and analyze the
structure of compact stars. It is most useful when one needs to
integrate Einstein equations numerically, but it is by no means the
only method.  In many instances Einstein equations can be simplified
due to symmetries or coincidences which are not displayed when use of
the TOV method is made. This is the case of some of the works
considered in this subsection.

By using Majumdar-Papapetrou matter, i.e., pressureless matter in
which the energy density and the electric charge density are equal in
appropriate units, and joining it into an extremal
Reissner-Nordstr\"om solution, Bonnor previously found that these
objects could have a radius as near as one wishes to their own
gravitational radius \cite{bonnor64,bonnor_wick} (see also
\cite{lemoszanchinonbonnor}). These star configurations are held
against collapse by electric repulsion.  Such solutions were improved
in \cite{lemosweinberg} where $C^\infty$ charged matter also showed
the transition to a new gravitational field state at the gravitational
radius of the configuration. This new gravitational state of a compact
star is called a quasiblack hole and is the realization of the
Buchdahl limit found in \cite{yu,andreasson_charged} for electric
configurations.  Quasiblack holes can also be considered the real
frozen stars, an alternative previous name given to black holes
\cite{novikzel}.

The inclusion of pressure into the matter also enabled to find exact
electric star solutions in which quasiblack holes, or frozen stars,
appear as limiting configurations \cite{lemoszanchin2010}.  Mention
should be made to the works \cite{defelice_yu,defelice_siming} where
consideration of an incompressible fluid in the presence of electric
charge, and using the TOV method, led to relativistic charged sphere
solutions which taking the limit to the black hole regime is still a
compact star, i.e., a quasiblack hole. Quasiblack holes also arise
from Yang-Mills--Higgs matter, Einstein-Cartan matter with spin and
torsion, rotating disk matter systems, and simple shells of matter,
see \cite{lemoskazan} for a review.

The properties of these very peculiar compact objects have been worked
out in detail, e.g., in \cite{lz1,lz5}, see also
\cite{meinel2011b}. They are objects on the limit of becoming extremal
black holes, but unlike a black hole, there is no collapse of the
progenitor star and the properties of matter inside the star are
relevant. The spacetime is regular throughout although falling
observers experience infinite tidal forces and the inner and outer
regions turn hermetic. For these compact objects one can find a mass
formula and their entropy yields results that match those for pure
black holes, see \cite{lemoskazan}.

\subsubsection {Compact stars with fields different from electric
and in alternative theories of gravity}
\label{sec-othertov}

There are general relativistic compact objects other that neutron
stars, black holes and quasiblack holes. An important class of such
objects are the regular black holes, i.e., objects that have all the
properties of black holes but do not show a singularity at their
core. For an electric realization of such objects see
\cite{lemoszanchin2011}.

Of course, fields other than Maxwell and charged matter can be put or
added, giving rise to several different compact objects such as
gravastars, boson stars and also regular black holes. In addition,
theories of gravitation, different from general relativity, can be
used giving rise to compact objects displaying their own peculiarities
showing that the study of compact stars comprises one of the
fundamental subjects in any gravitational theory, for a review see
\cite{lemospani}.

\subsection {This work}
\label{sec-thiswork}

Our aim here is to explore a particular class of spherically symmetric
cold charged fluid spheres, i.e., cold charged stars.  We investigate
electrically charged polytropic spheres.  Many features we have
mentioned will appear. The compact stars that will appear in this
study have an Oppenheimer-Volkoff limit, a Buchdahl limit and a
quasiblack hole limit.

The present paper is structured as follows. To properly define the
physical quantities of the model we write in
Sect.~\ref{sec-basicequations} the Einstein-Maxwell equations for a
charged perfect fluid. The explicit set of equations in the case of
spherical symmetry for a static spacetime in Schwarzschild-like
coordinates are given and put in the 
TOV form. To close the system, an
equation of state of polytropic form and a charge density profile are
defined.  Section \ref{sec-structure} is devoted to report the general
properties of the spheres in terms of the polytropic exponent.  The
Oppenheimer-Volkoff limit is analyzed as well as other properties of
the spheres, such as their radius to mass ratio. In taking an infinite
polytropic exponent, i.e., a stiff equation of state, one can find
with care the Buchdahl and quasiblack hole limits. A study of the
dependence of the speed of sound as a function of the polytropic index
is also performed.  In Section \ref{sec-conclusion} we conclude.

\section{Basic equations, the equations of equilibrium, and the
equations of state}\label{sec-basicequations}

\subsection{Basic equations}\label{sec-basicequationssub}

For completeness we start by writing the Einstein-Maxwell equations in
the presence of charged matter, $c=1=G$, \begin{eqnarray} &&
G_{\mu\nu}=8\pi T_{\mu\nu},\label{eqs de einstein}\\ &&
\nabla_{\nu}F^{\mu\nu}=4\pi J^{\mu}, \label{eqs de maxwell}
\end{eqnarray} where the Greek indices $\mu$, $\nu$, etc., run from
$0$ to $3$; the Einstein tensor is
$G_{\mu\nu}=R_{\mu\nu}-\frac{1}{2}g_{\mu\nu}R$, with $R_{\mu\nu}$ and
$g_{\mu\nu}$ being respectively the Ricci and the metric tensors, and
$R$ being the Ricci scalar.  $T_{\mu\nu}$ stands for the
energy-momentum tensor, which, in the present study is written as
$T_{\mu\nu}=M_{\mu\nu}+ E_{\mu\nu}$. $M_{\mu\nu}$ stands for the
energy-momentum tensor of a perfect fluid that is given by
\begin{equation}
M_{\mu\nu}=pg_{\mu\nu}+(p+\rho)U_{\mu}U_{\nu},
\end{equation}
where $\rho$ is the energy density, $p$ is the pressure, and $U_\mu$
is the fluid's four velocity.  The choice of a perfect fluid implies
that the flow of matter is adiabatic, no heat flow, radiation, or
viscosity is present \cite{misneresharp}. $E_{\mu\nu}$ is the
electromagnetic energy-momentum tensor,
\begin{equation}\label{tensor de energia momento}
4\pi
E_{\mu\nu}=F_{\mu}\hspace{0.1mm}^{\gamma}
F_{\nu\gamma}-\frac{1}{4}g_{\mu\nu}
F_{\gamma\beta}F^{\gamma\beta},
\end{equation}
where the Faraday-Maxwell strength tensor is
\begin{equation}
F_{\mu\nu}=\nabla_{\mu}A_{\nu}-\nabla_{\nu}A_{\mu},
\end{equation}
with $\nabla_\mu$ representing the covariant derivative, and ${\cal
A}_\mu$ the electromagnetic gauge field. In addition, the electric
current density is written as
\begin{equation}
J^{\mu}=\rho_{e}U^{\mu},
\end{equation}
where $\rho_{e}$ is the electric charge density.

\subsection{The equations of equilibrium}
\label{sec-equilibriumequations}

In order to describe a static fluid distribution with spherical symmetry,
the line element is assumed to be of the following form
\begin{equation}\label{geral_metric}
ds^2=-B(r)dt^{2}+A(r)dr^2+r^{2}d\theta^2+r^{2}\sin^{2}\theta
d\phi^{2} ,
\end{equation}
where $(t,\, r,\, \theta,\, \phi)$ are the usual Schwarzschild-like
coordinates, with the metric functions $A(r)$ and $B(r)$
depending on $r$ alone.

Imposing that there is a static spherically symmetric 
electric field  implies that the only nonzero
components of the Maxwell strength tensor 
is $F^{\mu\nu}$ are $F^{tr}=-F^{rt}$,
with $F^{tr}$ being a function of the radial coordinate $r$ alone, the other
terms of the Maxwell tensor are identically zero. Hence, the only
non-vanishing component of Maxwell equations (\ref{eqs de maxwell}) is
given by
\begin{equation}\label{continuidad da carga}
\frac{dq(r)}{dr}=4\pi\rho_{e}(r)\,r^{2} \sqrt{A(r)},
\end{equation}
where $q(r)=r^2\sqrt{A(r)\,B(r)}\,F^{tr}(r)$ is the total electric
charge inside a sphere of radial coordinate $r$, which does not depend
on the timelike coordinate $t$.

In the present case, considering the metric (\ref{geral_metric}), the
nonzero components of the Einstein equations (\ref{eqs de einstein})
are
\begin{eqnarray}
& &\hspace*{-.9cm} \frac{d}{dr}\left[r\, A^{-1}(r)\right]=1- 8\pi\,
r^2\left[\rho(r)
+\frac{q^{2}(r)}{8\pi r^{4}}\right], \label{G00final12}\\
& & \hspace*{-.9cm}
\frac{r\,B^{-1}(r)}{A(r)}\frac{dB(r)}{dr}+\frac{1}{A(r)}= {1}
+ 8\pi \,r^2\left[p(r)-\frac{q^{2}(r)}{8\pi r^{4}}\right].\label{G11final12}
\end{eqnarray}
As usual, we define a new quantity $m(r)$ representing the mass inside
the shell of radial coordinate $r$ in such a way that
\begin{equation}\label{funcion metrica}
A^{-1}(r)=1-\frac{2m(r)}{r}+\frac{q^{2}(r)}{r^{2}}.
\end{equation}
Now, replacing Eq.~(\ref{funcion metrica}) into Eq.~(\ref{G00final12}) it
gives
\begin{equation}\label{continuidad de la masa}
\frac{dm(r)}{dr}=4\pi\, r^2\rho(r)
+\frac{q(r)}{r}\,\frac{dq(r)}{dr},
\end{equation}
that represents the mass (energy) conservation, as measured in the
matter's frame.

Moreover, from the Bianchi identities $\nabla_{\mu}T^{\mu\nu}=0$, it
follows
\begin{equation}\label{conservacion2}
\frac{d B}{dr}=\frac{q\, B}{2\pi
r^{4}(p+\rho)}\left(\frac{dq}{dr}\right)-\frac{2\, B}
{p+\rho}\left(\frac{dp}{dr} \right),
\end{equation}
where, to shorten equations and simplify notation, we dropped the
function dependence, $B(r)=B$, $q(r)=q$, and so on.

Finally, replacing Eq.~(\ref{continuidad da carga}) and the
conservation equation (\ref{conservacion2}) into Eq.~(\ref{G11final12}) it
yields
\begin{equation}\label{tov}
\frac{dp}{dr}=-(p+\rho)\, \frac{\left(4\pi\,r\, p+\dfrac{m}{r^{2}}
-\dfrac{q^{2}}{r^{3}}\right)} {\left(1-\dfrac{2m}{r}
+\dfrac{q^{2}}{r^{2}}\right)}+\rho_{e}\sqrt{A\,}\,\frac{q}{r^{2} } ,
\end{equation}
which is the Tolman-Oppenheimer-Volkoff (TOV) equation
\cite{volkoff,tolman} modified to the study of equilibrium of an
electrically charged fluid, see \cite{bekenstein} (see also
\cite{ray}).

It is clear that equations (\ref{continuidad da carga}), (\ref{funcion
metrica}), (\ref{continuidad de la masa}) and (\ref{tov}) are not
enough to solve for the six variables $q(r)$, $A(r)$, $m(r)$,
$\rho(r)$, $p(r)$ and $\rho_{e}(r)$, since there are two degrees of
freedom.  The two missing equations (or constraints) are generally
obtained from a model of matter, in this case, of the charged
fluid. To complete this set of equations, it is commonly considered an
equation of state relating the pressure with the energy density of the
fluid. Moreover, for the electrically charged fluid, it is also needed
a relation defining the electric charge density, see below. The
resulting set of equations constitute the complete set of structure
equations which, with some appropriate boundary conditions, can be
solved simultaneously.

The final step to set up the system is defining the boundary
conditions for the sought solutions. In the present case, those are
chosen at $r=0$, the center of the sphere, and are: $m(r=0)=0$,
$q(r=0)=0$, $p(r=0) = p_c$, $\rho(r=0)=\rho_{c}$, $\rho_e(r=0) =
\rho_{ec}$. These conditions imply in $A(r=0)=1$. The surface of the
star $r = R$ is found when $p(R)=0$. For the numerical calculations,
the inputs in the system of equations are the central energy density
$\rho_{c}$ which through the equation of state yield the central
pressure $p_c$, and the central charge density $\rho_{ec}$.

The metric exterior to the sphere is given by Reissner-Nordstr\"om
metric
\begin{equation}\label{ext1}
ds^2 = - \left(1 -\frac{2M}{r} + \frac{Q^2}{r^2}\right) dT^2
+ \dfrac{dr^2} {1 -\frac{2M}{r} + \frac{Q^2}{r^2}}+ r^2d\Omega^2, 
\end{equation}
with $M$ and $Q$ being respectively the total mass and the total
charge of the sphere. The time $T$ is proportional to the inner time
$t$ and the radial coordinate $r$ is identical to the radial
coordinate of the interior region. 
The full set of boundary conditions at the surface of the star is
$B(R)=1/A(R)=1-2M/R+ Q^2/R^2$, $m(R)=M$, $q(R)=Q$, besides $p(R)=0$, 
and so $\rho(R)=0$ for a polytropic equation of state, which, 
together with a smooth charge density profile, is equivalent to
the full set of junction conditions for boundary layers in general
relativity in this particular instance.

\subsection{Equation of state and charge
density relation}\label{sec-densitychargerelation}

\subsubsection{The equation of state}
As mentioned above, to complete the set of equations it is necessary
to add two more relations to the charged fluid system. Usually, an
appropriate equation of state is furnished. Among the simplest
choices, a polytropic equation of state is frequently used. For the
purpose of the present analysis, following Tooper \cite{tooper1}, it
is convenient to choose the following polytropic relation
\begin{equation}
p = \omega \rho^\gamma\,, \label{eq_estado2}
\end{equation}
where $\omega$ is the polytropic constant and we use the name
polytropic exponent for the parameter $\gamma$.

Equation (\ref{eq_estado2}) is a good equation of state for a
Newtonian ideal fluid, indeed it is equivalent to that of a
non-relativistic fluid for small pressures, when compared to the energy
density \cite{htww}.  However, for high densities and pressures,
Eq.~\eqref{eq_estado2} violates causality conditions for any
$\gamma$. To see that we consider the speed of sound $c_{s}$ within
the fluid.  The square of the sound speed is $c^{2}_{s} =
{dp}/{d\rho}$ and so the equation of state (\ref{eq_estado2})
gives $c^{2}_{s}=\gamma\,{p}/{\rho}$.  On the basis of this
equality, we can determinate the value of the speed of sound across a
given star as a function of the radial coordinate $r$.  Moreover, we
can see that for sufficiently large pressure the speed of sound
becomes larger than the speed of light. In considering large values of
$\gamma$ here we will find high pressures at the center violating
causality and the usual energy conditions. Even though this is a
drawback, as we will see it is interesting to analyze the regime of
large polytropic exponents and check the properties of the
corresponding polytropic spheres. With this models in hand we can then
compare with other models of charged stars in the literature. We can
also verify how the speed of sound depends on the polytropic exponent,
central pressure, charge fraction and other physical quantities of the
model. Furthermore, in investigating the dependence of the structure
of the charged spheres on a gamut of different polytropic exponents
one can take the limit of very large $\gamma$ when the fluid becomes
incompressible and situations similar to the Schwarzschild interior
solution appear. More specifically, as we shall see bellow, in the
case of uncharged or charged spheres the Buchdahl limit is approached
for $\gamma\rightarrow \infty$. In addition, for very large $\gamma$
the quasiblack hole limit is reached.

\subsubsection{The charge density relation}

In order to study the effects of the electric charge in the structure
of polytropic charged spheres we need also to define the charge
density profile.  For simplicity, we assume that the charge density is
proportional to the energy density, see \cite{ray} (see also
\cite{zhang,siffert}),
\begin{equation}\label{densicarga_densimasa}
\rho_{e}=\alpha\rho,
\end{equation}
where $\alpha$ is the charge fraction and, in geometric units, is a
dimensionless constant.

As already mentioned, the effect of the electric charge in the
structure of the stars has been studied in previous works considering
several values of the charge fraction $\alpha$ for a fixed equation of
state, i.e., by taking Eq. (\ref{eq_estado2}) with fixed $\gamma =5/3$
\cite{ray}. In the present work we are interested in studying
all possible compact star solutions including their extreme limits,
such as quasiblack hole solutions. More specifically, using the
equation state \eqref{eq_estado2}, we consider different charge
amounts $\alpha$ and also different polytropic exponents
$\gamma$. Therefore, given the value of the charge fraction $\alpha$,
the polytropic constant $\omega$, and the central energy density (or the
central pressure), we can analyze the system as a function of
$\gamma$.

\section{The structure of polytropic charged stars with varying
$\bm\gamma$, $\bm{0<\gamma<\infty}$}
\label{sec-structure}

\subsection{General remarks}

We now present the structure of charged spheres made up of polytropic
fluids for various values of the polytropic exponent $\gamma$ and
charge fraction $\alpha$.  We investigate the type of compact objects
that might arise from these solutions, and whether they have a
quasiblack hole limit. In the analysis we have found that such a regime
is approached in the limit $\gamma\to \infty$.  In order to find the
whole spectrum of objects and in particular the quasiblack hole, we
integrate the system of equations considering many different values of
the polytropic exponent $\gamma$.  For the numerical integration we
rewrite equations (\ref{continuidad da carga}), (\ref{funcion
metrica}), (\ref{continuidad de la masa}), (\ref{tov}),
\eqref{eq_estado2}, and \eqref{densicarga_densimasa}, as well as the
boundary conditions adopted in the center of the star, in a
dimensionless form (see Appendix \ref{appendixA}). The relevant
equations then take the form given by Eqs.~(\ref{u}), (\ref{upsilon}),
and (\ref{theta}). For each given value of $\gamma$, $\alpha$, and
$\rho_c$ (or, equivalently, $p_c$), the resulting system of equations
is numerically solved through a fourth order Runge-Kutta method.  The
main results are displayed now.

Since we analyze the limit o large polytropic exponents, we plot
the results in terms of $\tanh \gamma$, for $\gamma$ in the interval
$\left[4/3, 17.1\right]$. Hence, it is interesting to show explicitly the
values of $\tanh\gamma$ as a function of $\gamma$. This is done in Table~I.
\begin{table}[h] \label{tabletanh}
\begin{tabular}{|c|c|c|c|c|c|c|c|c|}
$\gamma$& 4/3 & 5/3 & 2.0 & 3.0 &  5& 17.1 & $\infty$\\ \hline
$\tanh\gamma$& 0.8701 & 0.9311 & 0.9640 & 0.9950 & 0.9999 & 1.000 & 1.000
\end{tabular}
\caption{The hyperbolic tangent, $\tanh\gamma$, of the polytropic 
exponent $\gamma$.}
\end{table}

In presenting the numerical results we continue to put $c=1$ but $G$ is
given in  MKS units, $G= 6.67384 \times 10^{-11}\,{\rm m}^3/{\rm kg}\,
{\rm s}^{2}$.

\subsection{Numerical input values}

For a generic equation of state as \eqref{eq_estado2} in a star,
besides the polytropic exponent $\gamma$, there are two free
parameters, the polytropic constant $\omega$ and the central energy density
$\rho_c$. On the other hand, in order to analyze the structure of
charged polytropic spheres for large values of the polytropic exponent
one must be careful when choosing the normalization factors for the
numerical calculation, because the convergence strongly depends on
such a choice. Hence, for convenience, and also for the sake of
comparison with the results in the literature, the central energy density of
a standard neutron star model, that is, $\rho_c\sim 10^{17}\,{\rm
kg/m}^3$ (see, e.g., \cite{glendenning}), is a good reference value
for our numerical analysis.  With this in mind and recalling that the
equation of state \eqref{eq_estado2} is used, following \cite{ray}
(see also Appendix A) a value of $\omega$ is picked up in order that
for $\gamma=5/3$ and a suitably chosen central energy density, the solution
is close to the values for real neutron stars. In fact, it is convenient 
to normalize the polytropic constant $\omega$ in such a way that
it turns out a function of the polytropic exponent $\gamma$, namely,
$\omega=\omega(\gamma)=1.47518\times 10^{-3} \left(1.78266\times
10^{15}\,{\rm
kg}/{\rm m^3} \right)^{1-\gamma} $. With such a choice, a natural
normalization factor for the numerical calculations related to the TOV
equation would be ${\rho_0}= 1.78266\times 10^{15}{\, \rm kg/m}^3$,
corresponding to the pressure $p_0= 2.62974\times 10^{12} {\,\rm kg/m}^3$,
which is independent of $\gamma$. However, since we want to verify the limit
of arbitrarily large polytropic exponents (see below), it is necessary to
consider a normalization factor ${\rho_0}$ not larger than the central energy
densities of interest. We then have chosen to normalize quantities in terms
of the central energy density (see Appendix \ref{appendixA}).

We have made calculations for many different values of 
the polytropic exponent $\gamma$, the charge fraction $\alpha$, and the
central energy density $\rho_c$. We are mostly interested here in effects of
varying the exponent $\gamma$ in order to find, for high values of
$\gamma$, the quasiblack hole limit. We then fix the central energy density
and vary $\gamma$ starting from the value $\gamma=4/3$ which
corresponds to the usual relativistic ideal fluid. Note that for large
$\gamma$ the central pressure becomes very large,
see the equation of state  \eqref{eq_estado2}.

We show below the results obtained for different central energy
densities and some different values of the polytropic exponents.
Since one of the goals is the search
for quasiblack holes, we have chosen a central energy density ten
times larger than the normalization factor, $\rho_c=10\,{\rho_0}=
1.78266\times 10^{16}{\,\rm kg/m}^3$.  For this value of central energy
density, the maximum value of the polytropic exponent that produces
good results is approximately $17.0667$ ($17.1$, for short).  For
larger values, the convergence is slow and the results present
numerical instabilities.  In the case $\gamma = 17.1$ we have that the
central pressure is about $15.8$ orders of magnitude larger than the
 $\gamma =4/3$ central pressure, i.e.,
$p_c^{(\gamma=17.1)}/p_c^{(\gamma=4/3)}=
\omega^{(\gamma=17.1)}\rho_c^{17.1}/\omega^{(\gamma=4/3)}\rho_c^{4/3} 
\simeq 1.0\times 10^{15.8}$. Since $p_c^{(\gamma=17.1)}$ is much larger
than $p_c^{(\gamma=4/3)}$, in the numerical analysis we consider the first
value as satisfying the condition for the limit of arbitrarily large
pressure, we considered it as an infinite central pressure.

As usual, the values of the mass $M$, charge $Q$ and radius $R$ of the
star are found when the pressure at the surface of the star is equal
to zero $p(r=R)=0$. Since this rarely is the case, the numerical code
is stopped when the pressure becomes smaller than a appropriately
chosen very small value, or when it changes sign from positive to
negative values.

\subsection{Radius of the spheres as a function of the mass for fixed
polytropic exponent and different charge fractions: The
Oppenheimer-Volkoff limit}

\begin{figure}[!h]
\includegraphics[scale=1.12]{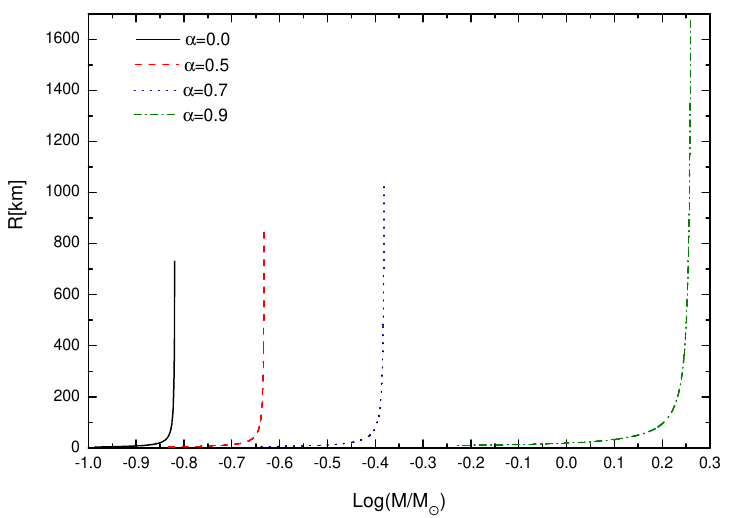}
\vspace*{-.5cm}
\caption{The radius of the charged polytropic sphere as a function
of the mass for $\gamma=4/3$ and a few values of
charge fraction $\alpha$. The considered central energy
densities are in the interval $1.0\times 10^{13}$ to $ 1.0\times10^{20}
{\,\rm  kg/m}^3$. The Chandrasekhar and Oppenheimer-Volkoff mass 
limits coincide in this $\gamma=4/3$ case.}
\label{R_vs_log_M43}
\end{figure}

\begin{figure}[!h]
\includegraphics[scale=1.12]{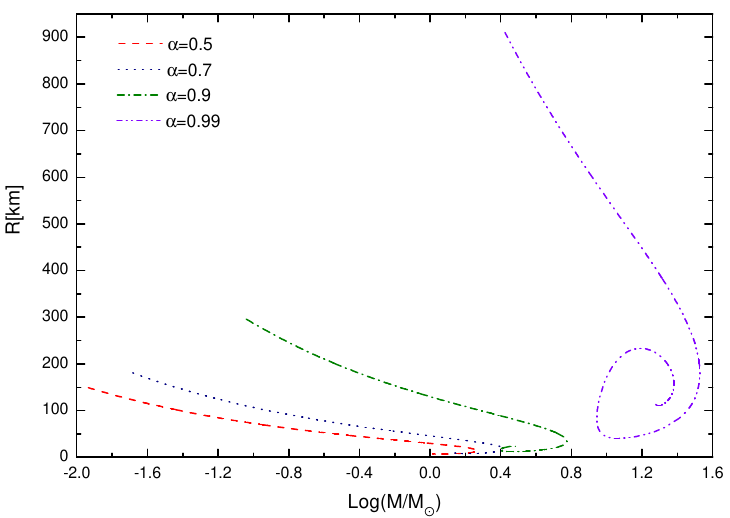}
\caption{The radius of the charged polytropic sphere as a function
of the mass for $\gamma=5/3$ and a few values of
charge fraction $\alpha$. The considered central energy
densities are in the interval $1.0\times 10^{13}$ to $ 1.0\times10^{20}
{\,\rm  kg/m}^3$. The Oppenheimer-Volkoff mass 
limit appears clearly, and there is no Chandrasekhar 
limit for $\gamma\neq4/3$.}
\label{R_vs_log_M}
\end{figure}

In Figs.~\ref{R_vs_log_M43} and
\ref{R_vs_log_M} we plot the radius of the
resulting spheres as a function of the mass,
normalized to the Sun's mass $M_\odot$, for $\gamma=4/3$ and
$\gamma=5/3$, respectively, and for a
few values of the charge fraction
$\alpha$. The exponent $\gamma=4/3$ represents soft relativistic
matter. Exponents like $\gamma=5/3$ or higher represent harder
matter cores.
The considered central energy densities $\rho_c$ are in the
interval $1.0\times 10^{13}$ to $ 1.0\times10^{20}\, {\rm
kg/m}^3$. These graphs are to be compared to the results of
Ref.~\cite{ray}, where $\gamma=5/3$ fixed and 
the maximum value of $\alpha$ is of about
$0.96$, while here we show results for $\alpha$ up to $0.99$. For a
given charge fraction, the usual behavior of the polytropic cold stars
is noticed, with the radius of the star decreasing with the mass as
the central energy density grows. Also, the spiraling behavior of the curve
$R\times M$ is observed for high central densities (see, e.g.,
\cite{mtw}).

In Fig.~\ref{R_vs_log_M43} for $\gamma=4/3$, one can see the
Chandrasekhar mass limit appearing distinctly in the vertical lines of the
plots together with the Oppenheimer-Volkoff limit, the points where
the vertical lines turn to the left.  Clearly, the two 
mass limits coincide
in this $\gamma=4/3$ instance, as perhaps 
could be expected following the heuristic arguments given
in \cite{landau}. Nonetheless it would be worth exploring this 
coincidence. In addition to a mass limit, the Oppenheimer-Volkoff
limit also gives a minimum radius for the star
(the Chandrasekhar minimum radius is zero, as it uses
Newtonian gravitation rather than general relativity).

In Fig.~\ref{R_vs_log_M} for $\gamma=5/3$, 
one can see the Oppenheimer-Volkoff limit,
the points where the inclined lines inflect to the left.  In this case,
as for all $\gamma\neq4/3$ cases, 
there is no Chandrasekhar limit.

\subsection{Properties of the spheres
as a function of the polytropic exponent: Towards the Buchdahl limit}

\subsubsection{Radius of the spheres
as a function of the polytropic exponent}

\begin{figure}[!h]
\centering
\includegraphics[scale=1.12]{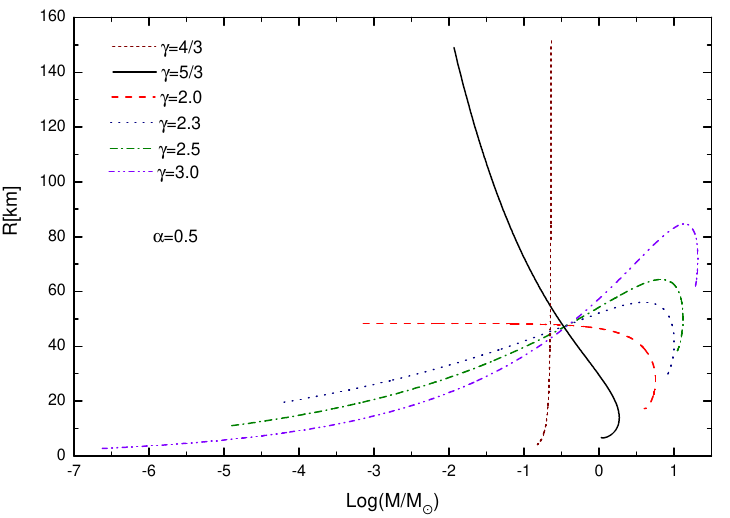}
\vspace*{-.5cm}
\centering
\includegraphics[scale=1.16]{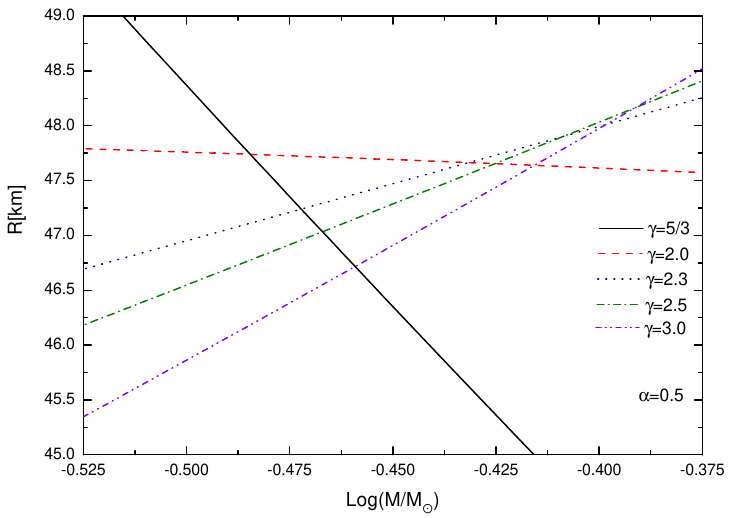}
\caption{Top: The radius of the charged polytropic sphere as a function
of the mass for $\alpha=0.5$ and a few values of the polytropic exponent
$\gamma$. The considered central energy
densities are in the interval $1.0\times 10^{13}$ to $ 1.0\times10^{20}
{\,\rm kg/m}^3$. The curve for $\gamma=4/3$ was interrupted at the
central energy density $\rho_c= 1.72\times10^{15}{\,\rm kg/m}^3$, the
complete behavior of this curve can be seen in the Fig. \ref{R_vs_log_M43}.
Bottom: Amplification of the region where the lines intersect. 
Note the lines do not intersect at a point as could be inferred
from the top panel. The line $\gamma=4/3$ is to the left of this 
bottom panel and does not appear.
}
\label{R_vs_log_M2}
\end{figure}

The dependence of the curve $R\times M$ on the polytropic exponent can be
seen in Fig.~\ref{R_vs_log_M2}, where we plot the radius of the
resulting spheres as a
function of the mass for $\alpha=0.5$ and a few values of the polytropic
exponent $\gamma$. As in the case of Fig.~\ref{R_vs_log_M}, the considered
central energy densities are in the interval $1.0\times 10^{13}$ to $
1.0\times10^{20}{\,\rm kg/m}^3$. These graphs show results completely new and
we do not find similar analysis in the literature for comparison. Namely, as
the polytropic exponent increases, the inclination of the curve $R\times M$
decreases becoming approximately horizontal for $\gamma$ around $2.0$. This
can be understood by taking into account that a very large polytropic
exponent implies an approximately constant energy density, $\rho = \rho_0$,
resulting in a relation of the form $M \sim 4\pi\rho_0 R^3/3$, which means
that the radius of the sphere increases with the mass and also with the
central energy density. It is interesting to say that the general behavior of
the curves for large $\gamma$ is independent of $\alpha$, namely the dominant
behavior of negative $dR/dM<0$ for small $\gamma$ (typically for $\gamma <
2.0$) changes to $dR/dM>0$ for large $\gamma$.

\begin{figure}[h]
\includegraphics[scale=1.12]{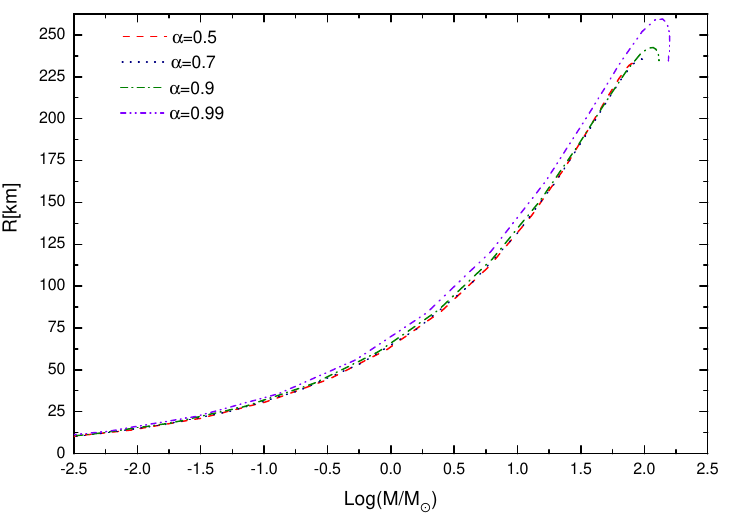}
\vspace*{-.5cm}
\caption{The radius of the charged polytropic sphere as a function
of the mass for $\gamma=17.0667$ and a few values of
charge fraction $\alpha$. The considered central energy
densities are in the interval $2.0\times 10^{15}$ to $ 2.4\times10^{16}
\,{\rm kg/m}^3$.} \label{R_vs_log_M3}
\end{figure}

We plot the radius of the resulting spheres as a function of the mass
for $\gamma=17.0667$ and a few values of the charge fraction $\alpha$ in
Fig.~\ref{R_vs_log_M3}. Notice that the charge fraction does not change
significantly the results. Since for large values of $\gamma$ the
central pressure becomes very large, the numerical calculations can not
be performed for large values of the central energy density, so that in this
case the considered values of $\rho_c$ are in the
interval $2.0\times 10^{15}$ to $ 2.4\times 10^{16}\,{\rm kg/m}^3$.

\subsubsection{Mass of the spheres as a function of 
the polytropic exponent}

\begin{figure}[!h]
\centering
\includegraphics[scale=1.16]{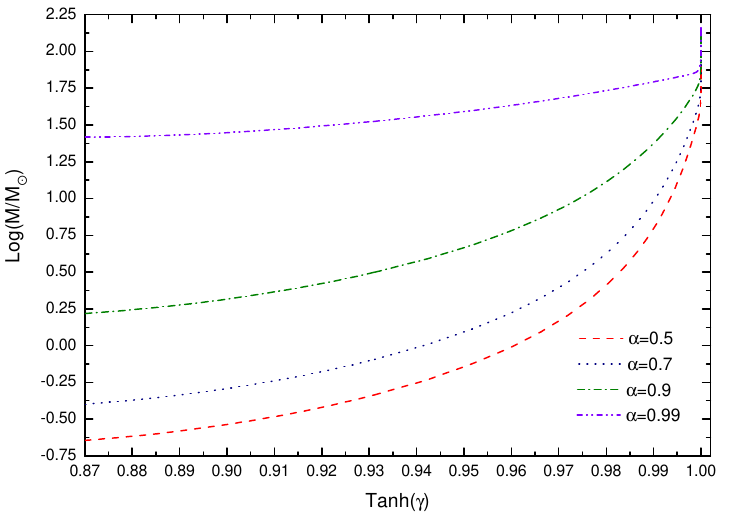}
\vspace*{-.5cm}
\caption{The mass of the charged polytropic sphere as a function of the
polytropic exponent considering the central energy
density $\rho_{c}=1.78266\times10^{16}\, {\rm kg/m}^{3}$ and a few values of
charge fraction $\alpha$.}
\label{log_gamma_vs_M}
\end{figure}

Fig. \ref{log_gamma_vs_M} shows the total gravitational mass of
the star as a function of the polytropic exponent (in logarithmic scale) for
the central density $\rho_c=1.78266\times10^{16}\, {\rm kg/m}^3$ and for
several values of the charge fraction $\alpha$. The values of $\gamma$
considered are in the interval $4/3 \leq\gamma\leq 17.0667$.
As one may observe, the gravitational mass increases with $\gamma$
in a rate that is larger for small charge fractions. For instance, in
the case $\alpha=0.5$, the mass for $\gamma =4/3$ is approximately $M =
0.23M_{\odot}$ while for $\gamma=17.1$ it is $M= 75.0M_\odot$. That is
to say, the mass grows about $32,509\%$. On the other hand, for
$\alpha= 0.99$ the mass grows approximately $588\%$. The growth of the
mass with $\gamma$ may be understood by noticing that the central
pressure increases with $\gamma$, so that the weight of more mass is
supported against collapse.

\begin{figure}[!h]
\includegraphics[scale=1.16]{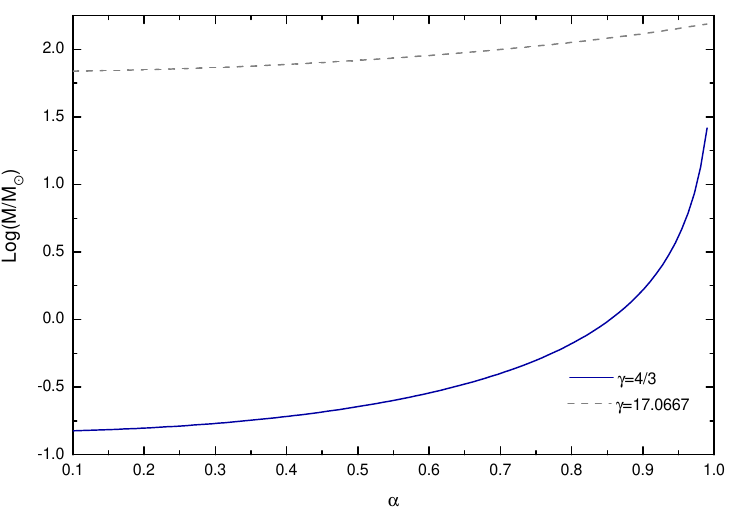}
\vspace*{-.5cm}
\caption{The mass of the charged polytropic sphere as a function of the
charge fraction $\alpha$ considering the central energy
density $\rho_{c}=1.78266\times10^{16}\, {\rm kg/m}^{3}$ and two values
of $\gamma$.}
\label{log_Mvsalpha}
\end{figure}

The dependence of the mass with $\alpha$ can also be seen in
Fig.~\ref{log_gamma_vs_M}. It increases monotonically with the charge
fraction, and strongly depends on $\gamma$ too. In fact, for $\gamma =
4/3$, the mass of the star increases about $11,451\%$, from $M = 0.23
M_\odot$ at $\alpha = 0.5$ to $M=26.2M_ \odot$ at $\alpha = 0.99$. In
turn, for $\gamma = 17.1$ and in the same interval of $\alpha$ the mass
varies approximately $85.57\%$, from $M=82.8M_\odot$ to
$M=153.65M_\odot$. This is shown explicitly in Fig.~\ref{log_Mvsalpha}.
Such a behavior is also easily explained since the electric charge acts
as an effective pressure, helping the hydrodynamic pressure to support 
collapse. Hence, the two effects of increasing $\alpha$ and $\gamma$ 
act favorable to yield equilibrium
configurations with large masses.

The value $\alpha=0.99$ is the maximum value of the charge fraction we
could find equilibrium solutions in our numerical calculations for the
largest value of $\gamma$ we considered. However, for $\gamma$ close to
$5/3$ we can get results even for $\alpha =0.9999$ corresponding to
spheres
with very large masses and radii.

\subsubsection{The radius to mass ratio
of the sphere as a function of the polytropic
exponent}

\begin{figure}[!h]
\includegraphics[scale=1.16]{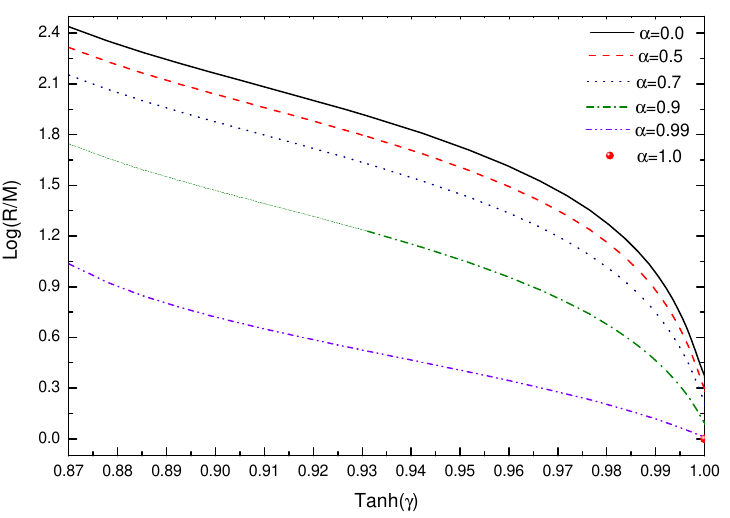}
\vspace*{-.5cm}
     \caption{Values of the ratio $R/M$ against the polytropic exponent
	$\gamma$ for a few values of the charge fraction $\alpha$. 
For large $\gamma$, the curves for $\alpha=0.0$ and for $\alpha=0.99$,
approach the Buchdahl limit $R/M=9/4$ and $R/M=1$, respectively.}
     \label{log_gamma_vs_R_M}
\end{figure}

\begin{figure}[!h]
     {
\includegraphics[scale=1.16]{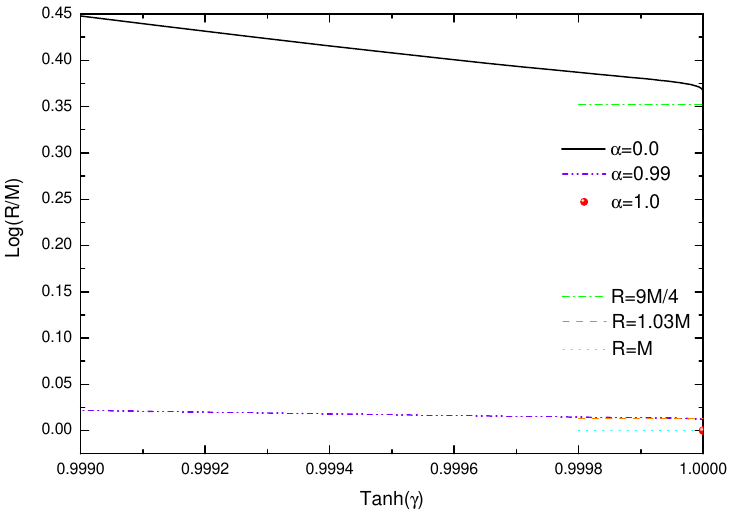}}
\vspace*{-.5cm}
\caption{The ratio $R/M$ as a function of the polytropic exponent
for $\rho_c=1.78266\times10^{16}\, {\rm kg/m} ^3$, $\alpha = 0.0$, and
$\alpha = 0.99$, showing the extreme limit in each case.
For $\alpha=0$ we get the Buchdhal limit, 
for $\alpha=1$ we get the quasiblack hole limit.}
     \label{R_M-of-gamma-limit2}
\end{figure}

The values of the ratio $R/M$ as a function 
of the polytropic exponent are shown
in Fig. \ref{log_gamma_vs_R_M} in logarithmic scale. In this figure, as in
Fig. \ref{log_gamma_vs_M}, it is considered the central energy density
$\rho_c=1.78266\times10^{16}\, {\rm kg/m^3}$ 
and a few different values of the charge
fraction $\alpha$. The main feature to note is that $R/M$ decreases with
increasing $\gamma$, a common behavior for all values of central energy
densities
we have investigated (even for those cases not showed in the figure). Note
that in the case without charge, $\alpha=0.0$, and large polytropic exponent,
$\gamma=17.1$, it is found $R/M \simeq 2.27$. 
If one extrapolates to $\gamma \to\infty$,
giving the Schwarzschild interior solution, 
one finds the Buchdahl
limit, $R/M=9/4=2.25$, as expected. This is shown more clearly in Fig.
\ref{R_M-of-gamma-limit2} (see also Sect. \ref{buchdahl-section}). On the
other hand, for
$\alpha = 0.99$ and $\gamma=17.1$ the ratio $R/M$ is very close to unity,
see the discussion below.

It is seen that for charged stars ($\alpha\neq 0.0$) we obtain
solutions that violate the original (uncharged) Buchdahl limit
\cite{buchdahl}, i.e., we find $R/M<9/4$. Also, there are values of $R/M$
that are smaller
than the bound for charged objects established in \cite{mak}, which
takes into account the contributions of the electric charge. For
instance, in the case $\alpha=0.99$ and $\gamma \simeq 17.1$, we find
$R/M\approx 1.03$, while the minimum value of $R/M$ calculated from
the results obtained in \cite{mak}, with our values for $M$ and $R$,
is $R/M \gtrsim 3.18$. A possible explanation for such a different
result is that our equation of state for large $\gamma$ implies
arbitrarily large values of the central pressure, while in \cite{mak}
it is assumed that the physical quantities are well behaved and
finite.

As seen here, the combined conditions of large charge fraction and
high polytropic exponent implies matter can be compressed beyond the
original Buchdahl limit, up to the quasiblack hole limit, see below.

\subsubsection{The charge of the sphere as a function of the polytropic
exponent}

\begin{figure}[!h]  
\includegraphics[scale=1.16]{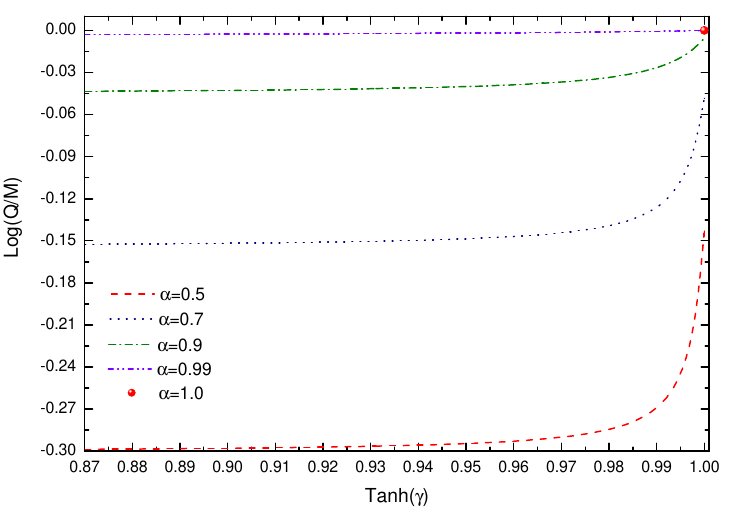}
\vspace*{-.5cm}
     \caption{Values of the ratio $Q/M$ as a function of the polytropic
exponent for different charge fractions $\alpha$ and central
energy density $\rho_{c}=1.78266\times10^{16}\, {\rm kg/m}^{3}$.}
     \label{log_gamma_vs_Q_M}
\end{figure}

The behavior of the charge of the sphere as a function of the
polytropic exponent can be seen in Fig. \ref{log_gamma_vs_Q_M}, where
we plot the ratio $Q/M$ versus the polytropic exponent in logarithmic
scale for the same central density, $\rho_c=1.78266\times10^{16}\,
{\rm kg/m}^3$, as in Fig.~\ref{log_gamma_vs_R_M}, and for different
values of charge fraction $\alpha$ as indicated. It is seen that $Q/M$
increases with the increasing of the polytropic exponent $\gamma$ for
all $\alpha$.  Moreover, the largest value $Q/M\approx0.999793$ is
found for the charge fraction $\alpha=0.99$ and
$\gamma=17.0667$. These are the largest values our numerical code
yield results without running into convergence troubles. For
intermediate values of $\alpha$, the results show an unexpected
dependence of the total amount of charge upon the polytropic
exponent. For a given charge fraction $\alpha$, the total charge jumps
from approximately $\alpha M$ for $\gamma\lesssim 2.0$ to
approximately $1.3\,\alpha M$ for $\gamma\gtrsim 4.0$. A possible
interpretation of this result is that fluid spheres made of stiff
matter (large $dp/d\rho$, see bellow) admit more charge than those
made of soft matter (small $dp/d\rho$).

\begin{figure}[!h]   
\includegraphics[scale=1.16]{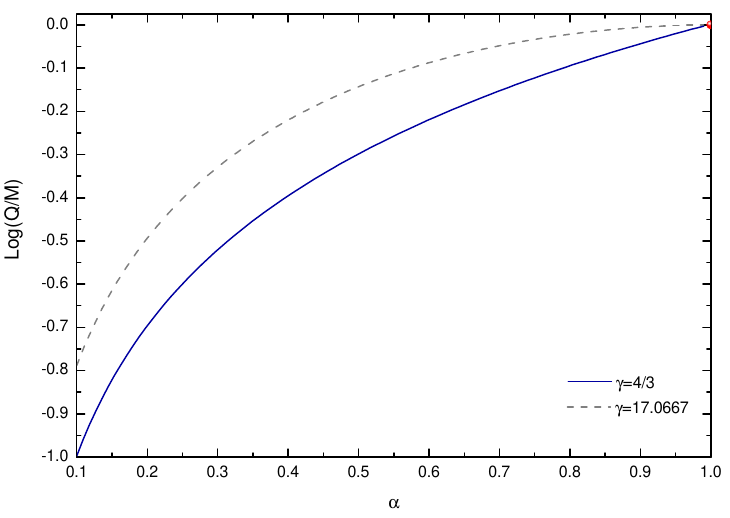}
\vspace*{-.5cm}
     \caption{Values of the ratio $Q/M$ as a function of the charge fraction
     for the polytropic exponent $\gamma=4/3$ and $\gamma=17.0667$ and
central energy density $\rho_{c}=1.78266\times10^{16}\, {\rm
kg/m}^{3}$. The curves for all $\gamma$ in the interval $[4/3,\;17.0667]$
are in between the two shown curves.}
     \label{Q_Mvsalpha}
\end{figure}

The dependence of the ratio $Q/M$ as a function of the charge fraction
$\alpha$ for some values of the polytropic exponent $\gamma$ is shown in
Fig.~\ref{Q_Mvsalpha}.


\subsection{The infinite polytropic exponent limit
and the Buchdahl and quasiblack hole limits}
\label{sec-large_gamma}

\subsubsection{Initial remarks}

The numerical analysis we have performed indicates that the quasiblack
hole limit is not reached by polytropic charged fluid spheres, using
the polytropic equation of state \eqref{eq_estado2}, with the
polytropic exponent in the interval $4/3\leq \gamma \leq 2.0$, and
with smooth boundary conditions, i.e., with finite central pressure
and zero pressure at the surface of the sphere.

However, previous results, and also some preliminary numerical
calculations using the present formulation, suggest that the
quasiblack hole limit can be found considering matter with equation of
state other than the polytropic one. For instance, in the case of
$\rho(r)= {\rm constant}$ \cite{defelice_yu,defelice_siming} or
constant total energy density \cite{lemoszanchin2010}, quasiblack
holes were modelled by charged matter. Moreover, in the present
framework we may get constant energy density in the limit of very
large polytropic exponents.

In order to investigate that limit normalized quantities are
necessary, as usual in numerical calculations. We call $p_{0}$ a
particularly chosen value of the central pressure and considered it as
the normalization factor for the pressure. Hence, using
Eq.~\eqref{eq_estado2} to write $p_{0} = w \rho_{0}^\gamma$, we get
\begin{equation}
\lim_{\gamma\rightarrow\infty}{\dfrac{p}{p_0}} =
\lim_{\gamma\rightarrow\infty }\left(\dfrac{\rho}{\rho_0}\right)^\gamma
              =\left\{\begin{array}{l}
		\infty, \quad {\rm if}\quad \rho > \rho_0;\\
		0, \quad {\rm if}\quad \rho < \rho_0\\
              \end{array} \right. \label{plimit}
\end{equation}
and
\begin{equation}
\lim_{\gamma\rightarrow\infty}{\dfrac{\rho}{\rho_0}} =
\lim_{\gamma\rightarrow\infty }\left(\dfrac{p}{p_0}\right)^{1/\gamma}
              = 1 . \label{rholimit}
\end{equation}
 The same holds for the charge density in case one uses
Eq.~\eqref{densicarga_densimasa},
\begin{equation}
\lim_{\gamma\rightarrow\infty}{\dfrac{\rho_e}{\rho_{0e}}} =
\lim_{\gamma\rightarrow\infty}{\dfrac{\rho}{\rho_0}}=
\lim_{\gamma\rightarrow\infty }\left(\dfrac{
p}{p_0}\right)^{1/\gamma}               = 1 .
\end{equation}
Therefore, if the central energy density $\rho_c$ is larger than the
normalization value $\rho_0$, the central pressure of such a sphere is
infinite.  On the other hand, if the central energy density is smaller
than $\rho_0$, the central pressure vanishes and then no equilibrium
solution is found.  Moreover, in such a limit, independently of the
value of $p_c$, i.e., independently of $\rho_c$, the energy density
$\rho$ is a constant throughout the charged fluid sphere. This is the
analogous of the Schwarzschild star, whose compactness is bounded by
the Buchdahl limit.

\begin{figure}[h!]  
\centering\includegraphics[scale=1.16]{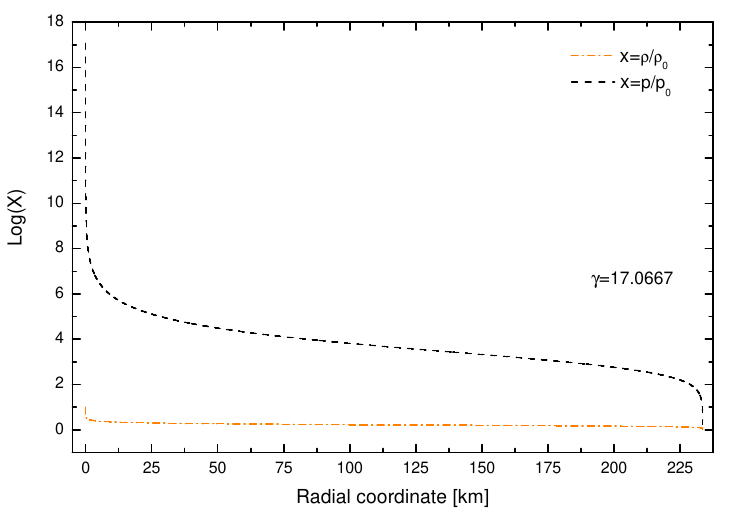}
 \vspace*{-.5cm}
\caption{The ratios $\rho(r)/\rho_0$ and $p(r)/p_0$ versus the radial
coordinate for the central energy density
$\rho_{c}=1.78266\times10^{16}\, {\rm kg/m}^{3}$,
polytropic exponent $\gamma=17.0667$ and charge fraction $\alpha=0.99$
The normalization factors used are  $\rho_{0}=1.78266\times10^{15}\,
{\rm kg/m}^3$ and $ p_0=2.62974\times10^{12}\, {\rm kg/m}^3$.}
\label{coordenada_radial_vs_rho_p}
\end{figure}

Fig.~\ref{coordenada_radial_vs_rho_p} shows the relations
$\rho(r)/\rho_0$ and $p(r)/p_0$ in terms of the radial coordinate for
the charge fraction $\alpha=0.99$, polytropic exponent
$\gamma=17.0667$ and central energy density
$\rho_{c}=1.78266\times10^{16}\, {\rm kg/m}^3$.  As it can be noted,
the pressure inside the sphere decreases very fast with radius.
Starting with a very high value at $r=0$, the pressure reaches its
minimum value, ideally equal to zero, at a value $R$ of the radial
coordinate, which is identified as the radius of the charged
sphere. Following the standard procedure, the numerical calculation is
then stopped at that point, $r = R$, and the interior solution is
matched to the exterior Reissner-Nordstr\"om solution. If the
calculation is continued, the pressure would reach negative values and
we discard those solutions. On the other hand, the energy density
varies very slowly with the radial coordinate, starting with
$\rho/\rho_0 =10$ at $r=0$ and decreasing only one order of magnitude
until very close to the surface of the sphere.  In comparison to the
pressure variation, the energy density is approximately a constant
throughout the sphere. It is important to mention that the initial
values of $\log\left(p/p_0\right)$ and $\log\left(\rho/\rho_0\right)$
differ by $16$ (sixteen), and that due to this the behavior of the
curve $\rho(r)/\rho_0$ is not fully clearly seen in
Fig.~\ref{coordenada_radial_vs_rho_p}.

With total confidence, we may then extrapolate these results and state
that our numerical analysis confirms the results for infinitely large
polytropic exponents, as shown by Eqs.~\eqref{plimit} and
\eqref{rholimit}. In such a limit the charged sphere is similar to the
Schwarzschild star in the sense that it has a constant energy
density. That peculiarity in the energy density of the star, such as
$\rho(r)= {\rm constant}$ like in \cite{defelice_yu,defelice_siming}
or total energy density equals a constant like in
\cite{lemoszanchin2010}, allows us to find quasiblack holes with
pressure using the hydrostatic equilibrium equation. Interestingly the
Schwarzschild interior solution has a constant energy density and,
numerically, the Buchdahl limit is attained by taking the limit of
infinite central pressure. Namely, the Schwarzschild stars satisfy
$R/M > 9/4$, with the upper bound for the most compact Schwarzschild
star, the Buchdahl limit $R/M = 9/4$, being found by taking the limit
$p_c \rightarrow \infty$ in the solution. We then expect to find
quasiblack holes in a similar situation, i.e, for static charged fluid
spheres with polytropic equation of state in the limit of very large
polytropic exponent. This was investigated numerically and the more
important results are presented in the next sections.

\subsubsection{The Buchdahl limit}
\label{buchdahl-section}

In order to see more clearly the extremely compact limit of the
objects studied in the present work we plot in
Fig. \ref{R_M-of-gamma-limit2} an amplified version of
Fig.~\ref{log_gamma_vs_R_M}, showing the limit of $R/M$ for large
$\gamma$. The aim of such a figure is to show the ratio $R/M$ for the
uncharged case ($\alpha=0.0$) and for the largest considered charge
fraction ($\alpha =0.99$) in the high polytropic exponent regime.

In the uncharged case, $\alpha=0.0$, 
the smallest value of ${R}/{M}$ we have found is
approximately $2.27$, which is very close to the Buchdahl bound,
${R}/{M}={9}/{4}=2.25$ \cite{buchdahl}.  The Schwarzschild interior
solution is probably the simplest case where the Buchdahl bound can be
verified. In fact, if the central pressure is allowed to be infinite
then the bound ${R}/{M}={9}/{4}$ 
is reached. In such a limit, the Schwarzschild
interior solution corresponds to an incompressible fluid (constant
energy density) with a monotonically decreasing pressure whose central
value is arbitrarily large.

In the charged $\alpha = 0.99$ case, 
the numerical result is also very close
to the extremal bound of 
Andr\'easson for the compactness of charged static spheres
\cite{andreasson_charged}.  In fact, from
Table~II one gets $R/M= 1.03$ and $R/R_+ =1.01$, indicating that the
radius of the charged sphere is really very close to its own
gravitational radius. This confirms the fact that the analogous to the
Buchdahl limit for charged static spherical objects is the quasiblack
hole limit, $R/M=R_+/M=Q/M=1$ (see below).

\begin{table*}
\begin{ruledtabular}
\begin{tabular}{ccccccccc}
$ $ & $\alpha$ & $M\times10^{5}\,[{\rm m}]$ & $Q\times10^{5}\,[{\rm m}]$ &
$R\times10^{5}\,[{\rm m}]$ & $R/M$ & $R/R_{+}$& $R/R_{-}$\\\hline
$1$ & $0.50$ & $1.22295$ & $0.879711$ & $2.32188$ & $1.89859$ & $1.12033$ &
$6.21802$\\
$2$ & $0.70$ & $1.47248$ & $1.31759$ & $2.33518$ & $1.58588$ & $1.09640$ &
$2.86491$\\
$3$ & $0.90$ & $1.92775$ & $1.90301$ & $2.34444$ & $1.21615$ & $1.04868$ &
$1.44728$\\
$4$ & $0.99$ & $2.27478$ & $2.27431$ & $2.33566$ & $1.02676$ &
$1.00631$ & $1.04807$
\end{tabular}
\caption{The obtained values of the mass, charge and
radius of the charged star, in geometric units, with the corresponding values
of $R/M$, $R/R_{+}$
and $R/R_{-}$, for several charge fractions $\alpha$ in the case
$ \rho_c=1.78266\times10^{16}\, {\rm kg/m}^3$ and $\gamma = 17.0667$.}
\end{ruledtabular}

\end{table*}

\subsubsection{The quasiblack hole limit}
\label{qbh-section}

As shown in Figs. \ref{log_gamma_vs_R_M} and \ref{log_gamma_vs_Q_M}, there
are values of $\alpha$ and $\gamma$ for which we find $R/M\approx1.02676$ and
$Q/M\approx0.999793$, in other words $R\approx M\approx Q$, indicating that
the quasiblack hole limit is about to be reached.
The best values we have found, i.e., the ones whose corresponding solution is
the closest to the quasiblack hole solution,
are obtained by considering the charge fraction $\alpha=0.99$ and
the polytropic exponent $\gamma=17.0667$. For higher values of $\alpha$ or
$\gamma$, our numerical code fails to converge. Therefore, the values
$\alpha=0.99$ and $\gamma=17.0667$ were chosen as best values and the other
functions and properties of the corresponding solution were determined.

Following \cite{lz1}, given a static spherically symmetric spacetime
solution  one has to check also the behavior of the metric functions in order
to decide if the solution is really a quasiblack hole, or, at least, really
close to one. For that we have studied the behavior of $A(r)$
and $B(r)$ for large $\gamma$ and large $\alpha$.

\begin{figure}[!h]
\centering\includegraphics[scale=1.16]{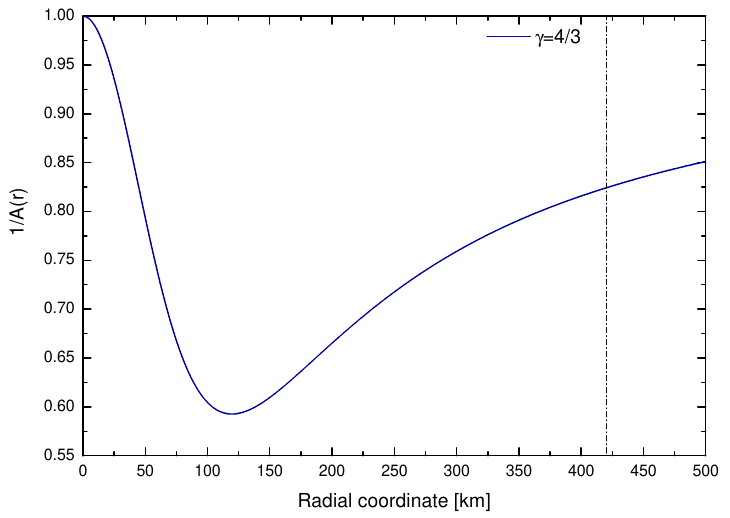}
\includegraphics[scale=1.16]{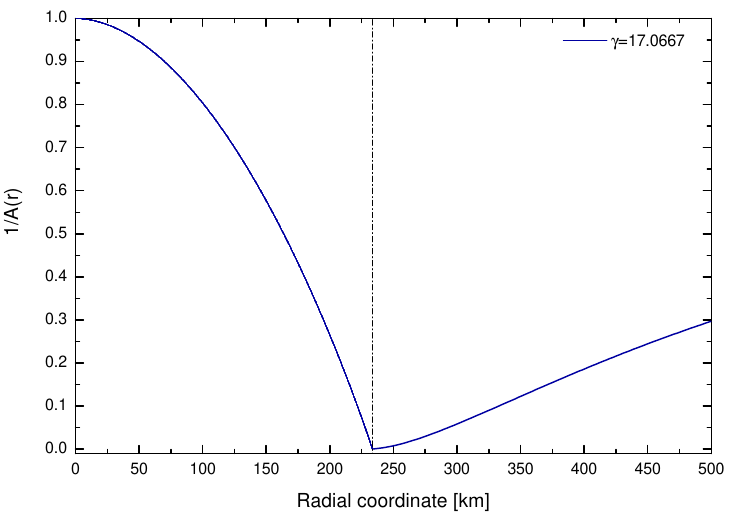}
\vspace*{-0.5cm}
     \caption{The metric function $A^{-1}(r)$ as a function
     of the radial coordinate for $\gamma=4/3$ (top) and
$\gamma=17.0667$ (bottom). In both cases $\alpha =0.99$. The 
vertical dashed line indicates the surface of the star.}
     \label{coordenada_radialvs1_grr}
\end{figure}

We plot the metric function $A^{-1}(r)$ against the radial coordinate in
Fig.~\ref{coordenada_radialvs1_grr}, 
for the charge fraction
$\alpha=0.99$, polytropic exponents $\gamma=4/3$ and
$\gamma=17.0667$,
and central density $\rho_c=10\rho_0= 1.78266\times10^{16}\,{\rm
kg/m}^3$.
Note that $A^{-1}(r)$
decreases monotonically with $r$, so that its minimum value is found at the
surface of the sphere. In the case considered here, the minimum value of
$A^{-1}(r)$ is $4.05648\times10^{-4}$, indicating that we are close to the
quasiblack hole limit for such a metric function, i.e.,
$A^{-1}(r=R)=\epsilon$ at the surface of the object, $r=R$.
 
\begin{figure}[!h]
     \centering
\includegraphics[scale=1.16]{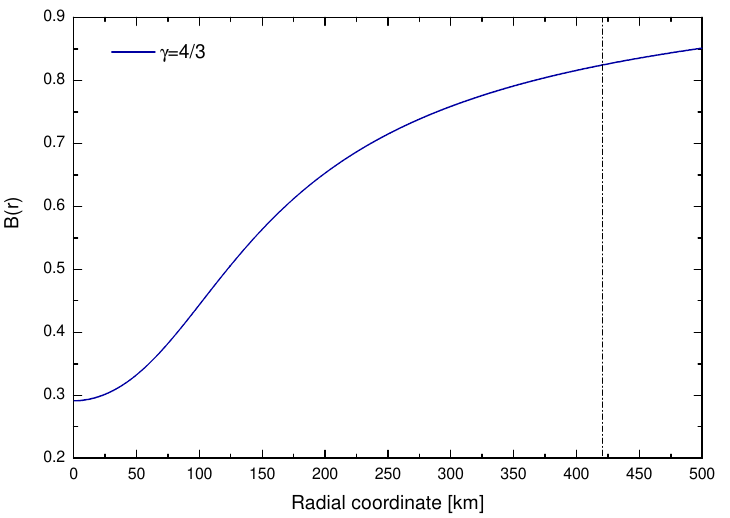}
\includegraphics[scale=1.16]{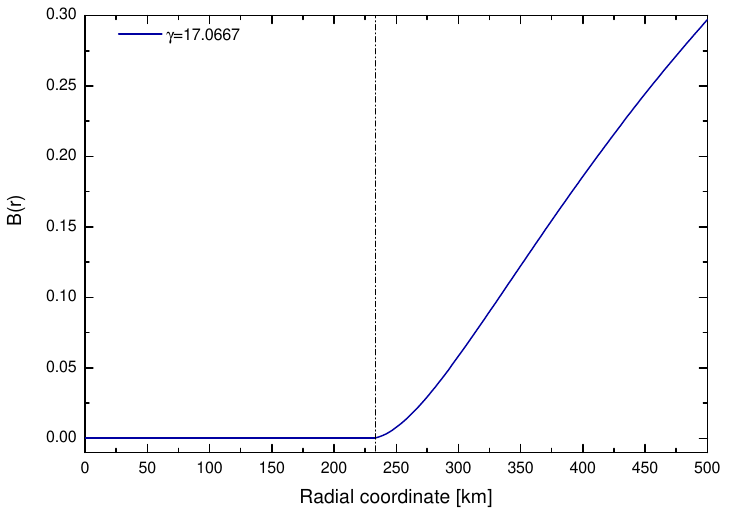}
\vspace*{-0.5cm}
     \caption{The metric function $B(r)$ as a function of the radial
coordinate  for $\gamma=4/3$ (top) and $\gamma=17.0667$ (bottom). In
both graphs we used $\alpha=0.99$. The dashed vertical line represents
the surface of the sphere.}
     \label{coordenada_radialvsgtt}
\end{figure}

In Fig.~\ref{coordenada_radialvsgtt} we plot the metric function
$B(r)$ as a function of the radial coordinate, for the charge fraction
$\alpha=0.99$, polytropic exponents $\gamma=4/3$ and $\gamma=17.0667$,
and central density $\rho_c=10\rho_0= 1.78266\times10^{16}\,{\rm
kg/m}^3$. Once the values of $M$, $Q$ and $R$ are already known, this
graphic is drawn by solving the conservation equation
(\ref{conservacion2}), integrating from the surface until the center
of the sphere. For these particular values of parameters, the metric
function $B(r)$ assumes very small values, increasing slowly with the
radial coordinate, to reach the largest value at the surface of the
star, where it equals the metric coefficient $A^{-1}(r)$. The maximum
value is $B(r)=4.05648\times10^{-4}$ at $r=R$. As just said, this
value is the same as the value of $A^{-1}(R)$, because at the surface
of the star one has
$\displaystyle{B(R)=A^{-1}(R)=1-\frac{2M}{R}+\frac{Q^2}{R^2}}$, which
is due to the fact that interior solution is matched smoothly to the
Reissner-Nordstr\"om exterior solution. It is worth noticing that the
metric function $B(r)$ is very small all along the interior of the
sphere, also indicating that the solution is close to the quasiblack
hole limit in which $B(r)$ is vanishingly small, $B(r) \rightarrow
\epsilon $ across the hole region inside matter.

The ratio $Q/M$ as a function of the charge fraction $\alpha$, for the
polytropic exponent $\gamma=17.0667$, and for the central density
$\rho_c=1.78266\times10^{16}\,{\rm kg/m}^3$ has been shown in Fig.
\ref{log_gamma_vs_Q_M}. As can be seen in that figure, the ratio $Q/M$
approaches one with the increasing charge fraction. For $\alpha=0.99$,
the values of the charge and of the mass of the given star are very
close to each other $(Q\approx M)$, which implies that $R_\pm\equiv
M\pm\sqrt{M^{2}-Q^{2}} \approx M$, where $R_\pm$ are, respectively,
would be event horizon and the Cauchy horizon of the corresponding
star, found as usual through the solutions of the equation $B(R)=0$.
Moreover, from the values of ratios $Q/M$ and $R/M$, for the charge
fraction $\alpha=0.99$ and $\gamma=17.0667$, we conclude that the
radius of the charged sphere is larger than the radius of the event
horizon of the corresponding Reissner-Nordstr\"om black hole, $R_+$,
indicating that we have a static equilibrium configuration.  Also,
since $R/R_{+}\approx 1$, the boundary of the star approaches its own
gravitational radius, which, together with the other properties found
above, indicates that the quasiblack holes limit is about to be
reached, see also \cite{defelice_yu,defelice_siming} and
\cite{lemoszanchin2010}.

The values of $M$, $Q$, $R$, $R_\pm$ and their relations are shown in
Table~II for the maximum $\alpha$ and $\gamma$ that permit good
numerical results, namely, $\alpha=0.99$ and $\gamma=17.0667$. We
obtain $Q \approx M$, in geometric units, and then $R\simeq R_+\simeq
R_-$. These results, together with the fact that $B(r)\sim\epsilon $
for all $0\leq r\leq R$ and $A^{-1}(R) \sim\epsilon$ (small
$\epsilon$) guarantee that we are close to the quasiblack hole limit
(see \cite{lz1} for a precise definition a quasiblack hole).

As well known, similar extreme relations characterizing quasiblack
holes, such as $R_{+}\simeq R_{-}\simeq M=Q$, are found in the case of
charged dust stars, with zero interior pressure ($p=0$). This follows
mainly in charged systems that satisfy the Majumdar-Papapetrou
conditions, see, e.g, \cite{lemoskazan}.  The present result is an
additional example of quasiblack holes with pressure, as the one found
in Ref.~\cite{lemoszanchin2010} (see also \cite{lz5}).

\subsection{Speed of sound within the fluid and the 
causality condition}\label{sec-sound_speed}

If one is interested in restricting the charged sphere solutions to those
that do not violate causality, the speed of sound inside the fluid is an
important property to investigate.

The speed of sound is defined through 
the equation $c^{2}_{s} = {dp}/{d\rho}$. So from
the equation of state (\ref{eq_estado2}) one finds
\begin{equation}\label{sound_speed2}
c^{2}_{s} = \frac{dp}{d\rho}=
\gamma\,\frac{p}{\rho}=\omega\gamma\rho^{\gamma-1}\,.
\end{equation}
As seen from Eq.~\eqref{sound_speed2}, the speed of sound 
gets larger than
the speed of light if
\begin{equation}\label{sound_speed2a}
  \omega\gamma\rho^{\gamma-1}> 1.
\end{equation}
For the charged 
polytropic fluid we are considering here, the energy density
$\rho(r)$ decreases towards the surface of the star and so does the
speed of sound. This is true for finite $\gamma$. In other words, for
a given equilibrium solution, the speed of sound is greatest at the
center of the star. This has been confirmed numerically for all values
of the parameters we have checked. Hence to test the causality
condition, $c_s\leq 1$, it is sufficient to determine the speed of
sound at the center of the star.

\begin{figure}[!h]
     \centering
     \includegraphics[scale=1.16]{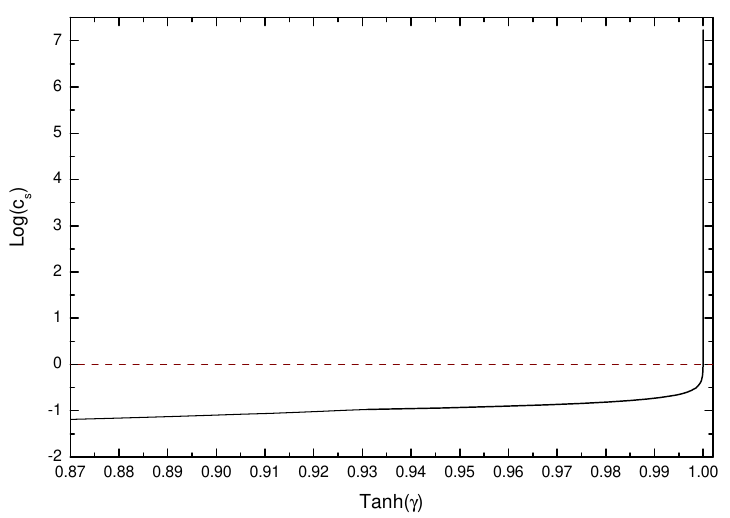}
\vspace*{-0.5cm}
     \caption{Speed of sound in the center of the star against the
    polytropic exponent $\gamma$ for the central density
     $\rho_c=1.78266\times10^{16}\,{\rm kg/m}^3$. For $\gamma \gtrsim 3.3$
the speed of sound $c_s$ is larger than the speed of light.}
     \label{gamma_vs_som}
\end{figure}

It is clear from Eq.~\eqref{sound_speed2} that, for relatively small
energy densities and small
polytropic exponents, $c_s^2$ is smaller than
unity. However, as the polytropic index grows, the fluid gets stiffer
and eventually becomes incompressible for
$\gamma\rightarrow\infty$. Hence, for given values of the polytropic
constant $\omega$ and central energy density $\rho_c$, there is a particular
value of $\gamma$ above which causality is violated. In the case of
Fig.~\ref{gamma_vs_som} it is $\gamma\simeq 3.3$.

In the graphic of Fig. \ref{gamma_vs_som} we plot the speed of sound
at the center of the star $(c_{s})$ as a function of the polytropic
exponent $(\gamma)$ for the central energy density
$\rho_c=1.78266\times10^{16}\, {\rm kg/m}^3$. It is seen that for
large values of $\gamma$ the speed of sound exceeds the speed of
light. This is because as $\gamma$ grows very large, the value of
$\rho(r)$ approaches a constant and $dp/d\rho$ becomes arbitrarily
large.  The speed of sound strongly depends on the central energy density,
and increases with the exponent $\gamma$. In the limit of very large
$\gamma$ the charged spheres become incompressible for which the sound
speed cannot be defined by Eq.~\eqref{sound_speed2}. In such a limit,
compact uncharged spheres can be found, these being
the interior
Schwarzschild solution with the most compact one given by the
Buchdahl limit. In the 
charged case the limiting $\gamma\to\infty$ solution 
is again a $\rho(r)= {\rm constant}$ solution and it yields
the quasiblack hole limit.

\section{Conclusions}
\label{sec-conclusion}

We have studied electrically charged polytropic spheres in the context
of Einstein-Maxwell theory. The spheres contain a spherically
symmetric distribution of charged perfect fluid, and the exterior
spacetime is represented by the Reissner-Nordstr\"om metric.  The
charge density $\rho_{e}$ and the energy density $\rho$ were assumed to
have the relation $\rho_{e}=\alpha\rho$, whereas the fluid assumes a
polytropic equation of state relating the pressure $p$ and the energy
density $\rho$ of the fluid, $p=\omega\rho^{\gamma}$, with $\omega$ and
$\gamma$ being constants. The choice of parameter $\omega$ is such that, for
$\gamma=5/3$ and appropriate central energy density $\rho_c$, the system is
close to the realistic neutron stars.  We have studied the
Oppenheimer-Volkoff limit, the Buchdahl limit and the quasiblack hole
limit.

We have analyzed the configurations for several $\gamma$, from
$\gamma=4/3$ to $\gamma$ reasonably high. Indeed, we have found that
for a central energy density that is ten times larger than the normalization
factor the highest value of the polytropic exponent that produced
proper numerical results is $\gamma=17.0667$. In such a case, the
numerical results approached interesting limiting cases.

In the zero charge case, $\alpha=0.0$, for all of the central energy
densities considered, in the limit of large polytropic exponents we
found solutions for stars that are very close to the Buchdahl
limit, $R/M = 9/4$. The mass to radius relation increases with the
polytropic exponent attaining a value very close to $9/4$ for $\gamma
= 17.0667$. In such a limit the spheres are similar to the
Schwarzschild star since they have a constant energy density, the central
pressure in this limit being arbitrarily large.

For the charged case, for fixed finite $\gamma$, and varying the
central pressure we have not found quasiblack holes.  On the other
hand, in the limit of very high $\gamma$, and with the central
pressure tending to infinity, we have shown that the quasiblack hole
limit is reached.  In fact, with increasing polytropic exponent and
charge fraction the relation $R/M$ approaches unity. The largest value
of such a ratio is found considering $\alpha=0.99$ and
$\gamma=17.0667$, which are the highest values our numerical code
furnished trustworthy results. For these values of $\alpha$ and
$\gamma$ we found also that $Q\simeq M$, and so the radius of the
sphere is close to the corresponding horizon radius $R\simeq R_{+}= M
+\sqrt{M^2 - Q^2}$. In addition to this we verified other properties
of the charged spheres that indicate beyond doubt the presence of
quasiblack holes with pressure.

The physical properties of charged spheres with other choices of
equations of state and other charge density profiles should also be
investigated, we shall report such analyses in a future work.

\begin{acknowledgments}
JDVA thanks Funda\c{c}\~ao Universidade Federal do ABC - UFABC, Brazil,
and Coordena\c{c}\~ao de Aperfei\c{c}oamento de Pessoal de N\'\i vel
Superior - CAPES, Brazil, for a grant.  VTZ would like to thank
Conselho Nacional de Desenvolvimento Cient\'ifico e Tecnol\'ogico -
CNPq, Brazil, for grants, and Funda\c{c}\~ao de Amparo \`a Pesquisa do
Estado de S\~ao Paulo for a grant (Processo 2012/08041-5).  JPSL
thanks the support of the Funda\c{c}\~{a}o para a Ci\^{e}ncia e a
Tecnologia of Portugal - FCT, projects PTDC/FIS/098962/2008 and
PEst-OE/FIS/UI0099/2011. JPSL and VTZ thank
the Observat\'orio Nacional do Rio de Janeiro - ON, Brazil, for
hospitality.
\end{acknowledgments}

\appendix
\section{Dimensionless relativistic equations of a
polytrope}\label{appendixA}

For the numerical calculations, the relativistic equations of a polytrope
must be written  in
dimensionless form. For this, we introduce the dimensionless
radial coordinate $\varepsilon$ given by 
\begin{equation}
\varepsilon= r{\sqrt{4\pi\rho_c}}, \label{normradius}
\end{equation}
and the new variables $\upsilon(\varepsilon)$, $u(\varepsilon)$, and 
$\theta(\varepsilon)$ defined by
\begin{eqnarray}
\upsilon(\varepsilon)= {\sqrt{4\pi\rho_c}}\,m(r)\label{normmass},\\
u(\varepsilon)= \frac{\sqrt{4\pi\rho_c}}{\varepsilon^2}\,q(r),
\label{normcharge}\\
\theta(\varepsilon) = \left(\frac{\rho(r)}{\rho_c}\right)^{\!\gamma},
\label{normdensity}
\end{eqnarray}
$\rho_{c}$ being the central energy density of the star. In terms of the
normalized energy density $\theta$, the pressure becomes 
$p(r)= \omega{\rho_{c}}^{\gamma}\,\theta(\varepsilon)$. Considering
the dimensionless variables $u$, $\upsilon$, $\theta$, $B$ and 
the relation
(\ref{densicarga_densimasa}), the relativistic equations of a
polytrope, Eqs.~(\ref{continuidad da carga}), (\ref{continuidad de la
masa}), and (\ref{tov}) in a dimensionless form are
\begin{eqnarray}
\label{u}
& &\frac{du}{d\varepsilon}= -\frac{2u}{\varepsilon}+
\dfrac{\alpha\theta^{1/\gamma}}
{\sqrt {1- \dfrac{2\upsilon}{\varepsilon} + \varepsilon^2 u^2 }}, \\
\label{upsilon}
& &\frac{d\upsilon}{d\varepsilon}= \theta^{1/\gamma}\varepsilon^{2}
+ \dfrac{ \alpha\varepsilon^3 \theta^{1/\gamma} u}
{ \sqrt{ 1- \dfrac{2\upsilon}{\varepsilon} + \varepsilon^2u^2\,}},\\
\label{theta}
& &\frac{d\theta}{d\varepsilon}=
-\varepsilon\left(\theta+\frac{{\rho_c}^{1-\gamma}}{\omega}\theta^{1/\gamma }
\right)\left(
  \dfrac{ \displaystyle{
 \omega{\rho_c}^{\gamma-1}\theta- u^2 +
\frac{\upsilon}{\varepsilon^3} }} {1- \dfrac{2\upsilon}{\varepsilon}
+\varepsilon^2 u^2}\right) \nonumber \\ & &\hskip 1cm+
\dfrac{\displaystyle{\alpha\,{\rho_c}^{1-\gamma}{\omega^{-1}}
u\theta^{1/\gamma}}} { \sqrt{1- \dfrac{2\upsilon}{\varepsilon} +
\varepsilon^2 u^2 \,}}.
\end{eqnarray}
This gives a set of three coupled differential equations,
(\ref{u})--(\ref{theta}), that are solved simultaneously to get the
equilibrium solutions. The boundary conditions adopted in the center
of the star, where $\varepsilon=0$, are $\upsilon(0)=0$, $u(0)=0$ and
$\theta(0)=1$. The maximum value of $\varepsilon$ is found
when $\theta(\varepsilon)=0$, and such a particular value of 
$\varepsilon$ is identified as the radius of the polytropic sphere,
$\varepsilon= \varepsilon_s$,

In astrophysics of neutron stars, units appropriate to nuclear physics
\cite{glendenning} are usually used, so that, for units in which the
speed of light is set to unity ($c=1$), the pressure $p$ and energy
density $\rho$ are given in MeV/fm$^{3}$. Moreover, from the relation
(\ref{densicarga_densimasa}), the charge density $\rho_{e}$ is also
measured in MeV/fm$^{3}$. The reference energy density, $\rho_0$, used
as a normalization factor in the numerical calculations, is
$\rho_0 = 1.0\,$ MeV/fm$^{3}$. Transforming to MKS units,
which we use in this paper, this means
$\rho_0=1.78266\times10^{15}\,{\rm kg/m}^3$. The normalization
factor may be changed according to the region of the central energy density
one is interested, or depending on the equation of state of the fluid
being used. We have done both variations, but in this paper we report
the results obtained by using the TOV equation normalized in terms of the
central energy density $\rho_c$, as given by Eqs.~\eqref{normradius},
\eqref{normmass}, \eqref{normcharge} and \eqref{normdensity}.

In the present analysis, the polytropic constant $\omega$ is normalized in
terms of the reference central energy density $\rho_0 = 1.0\,$ MeV/fm$^{3}$.
The particular value $\omega=1.47518\times10^{-3}\left[{\rm
fm}^{3}/\,{\rm MeV}
\right]^{\gamma-1}$, that is equivalent to $\omega=1.47518\times
10^{-3} \left(1.78266\times 10^{15}
\right)^{1-\gamma}\left[\dfrac{{\rm m}^3}{\rm kg}\right]^{\gamma-1}$,
was chosen for the sake of comparison to the results of previous works
\cite{ray}. Notice that, with this choice, the parameter $\omega$ results to
be a function of the polytropic exponent, $\omega= \omega(\gamma)$.

The integration of equations \eqref{u}, \eqref{upsilon} and
\eqref{theta} is stopped at the point $\varepsilon$ 
where the pressure $\theta(\varepsilon)$ reaches negative values, or
otherwise, when it gets
smaller than an appropriate chosen value, $\theta(\varepsilon)\sim 0$. The
corresponding vale of the radial coordinate
$\varepsilon=\varepsilon_s$ is extracted and the radius of the sphere
is obtained from the relation
$R=\dfrac{\varepsilon_s}{\sqrt{4\pi\rho_c}}$. Then, the physical
quantities, mass $M$ and charge $Q$, are calculated respectively from
$M\equiv m(R)=\dfrac{\upsilon(\varepsilon_s)}{\sqrt{4\pi\rho_c}} $ and
$Q\equiv
q(R)=\dfrac{\varepsilon_s^{2}u(\varepsilon_s)}{\sqrt{4\pi\rho_c}} $.

After obtaining $\upsilon(\varepsilon)$ and $u(\varepsilon)$ the metric
functions $B(\varepsilon)$ and $A(\varepsilon)$ are determined
from the relations
\begin{eqnarray}
\label{B1}
& &\hspace{-.8cm} \frac{dB(\varepsilon)}{d\varepsilon}=2\varepsilon\,
B(\varepsilon)\left(
\dfrac{\displaystyle{ \omega\,{\rho_c}^{\gamma-1}\,\theta(\varepsilon)-
u^2(\varepsilon)+
\frac{\upsilon(\varepsilon)}{\varepsilon^3}}} {1-
\dfrac{2\upsilon(\varepsilon)}{\varepsilon} +
\varepsilon^2 u^2(\varepsilon)}\right), \\
\label{A1}
& & \hspace{-.8cm} A^{-1}(\varepsilon) = 1-
\dfrac{2\upsilon(\varepsilon)}{\varepsilon} +
\varepsilon^2 u^2(\varepsilon).
\end{eqnarray}

\end{document}